\documentclass[10pt,journal]{IEEEtran}

\usepackage[pdftex]{graphicx}
\usepackage{algorithm}
\usepackage{color}
\usepackage{caption}
\usepackage{algpseudocode}

\definecolor{ForestGreen}{RGB}{162,52,0}

\usepackage{xcolor}
\usepackage{amsmath}
\usepackage{CJK}
\usepackage{url}
\usepackage{array}
\usepackage{ragged2e}

\ifCLASSOPTIONcompsoc
  \usepackage[nocompress]{cite}
\else
  \usepackage{cite}
\fi

\ifCLASSINFOpdf
\else
\fi

\ifCLASSOPTIONcompsoc
  \usepackage[caption=false,font=footnotesize,labelfont=sf,textfont=sf]{subfig}
\else
  \usepackage[caption=false,font=footnotesize]{subfig}
\fi

\begin{document}

\title{IoV Scenario: Implementation of a Bandwidth Aware Algorithm in Wireless Network Communication Mode}

\author{Peiying Zhang,
Chao Wang,
Gagangeet Singh Aujla,~\IEEEmembership{Senior Member,~IEEE},
Neeraj Kumar,\\~\IEEEmembership{Senior Member,~IEEE} and
Mohsen Guizani,~\IEEEmembership{Fellow,~IEEE}

\thanks{Manuscript received XX, XX, 2020; revised XX, XX, 2020. This work is partially supported by the Major Scientific and Technological Projects of CNPC under Grant ZD2019-183-006, and partially supported by ``the Fundamental Research Funds for the Central Universities" of China University of Petroleum (East China) under Grant 20CX05017A, 18CX02139A. \textit{(Corresponding author: Neeraj Kumar.)}}
\thanks{Peiying Zhang and Chao Wang are with the College of Computer \& Communication Engineering, China University of Petroleum (East China), Qingdao 266580, P. R. China. E-mail: 25640521@qq.com.}
\thanks{Gagangeet Singh Aujla is with School of Computing, Newcastle University, Newcastle Upon Tyne, United Kingdom and Computer Science and Engineering Department, Chandigarh University, Gharuan, Mohali, Punjab, India. E-mail: gagi\_aujla82@yahoo.com.}
\thanks{Neeraj Kumar is with Computer Science \& Engineering Department, Thapar Institute of Engineering and Technology (Deemed to be University), Patiala, India. Emails: neeraj.kumar@thapar.edu.}
\thanks{Mohsen Guizani is with CSE Department, Qatar University, Doha 2713, Qatar. E-mail: mguizani@ieee.org.}
}

\markboth{IEEE TRANSACTIONS ON VEHICULAR TECHNOLOGY,~Vol.~XX, No.~XX, XX~2020}
{}

\maketitle
\begin{abstract}
The wireless network communication mode represented by the Internet of vehicles (IoV) has been widely used. However, due to the limitations of traditional network architecture, resource scheduling in wireless network environment is still facing great challenges. This paper focuses on the allocation of bandwidth resources in the virtual network environment. This paper proposes a bandwidth aware multi domain virtual network embedding algorithm (BA-VNE). The algorithm is mainly aimed at the problem that users need a lot of bandwidth in wireless communication mode, and solves the problem of bandwidth resource allocation from the perspective of virtual network embedding (VNE). In order to improve the performance of the algorithm, we introduce particle swarm optimization (PSO) algorithm to optimize the performance of the algorithm. In order to verify the effectiveness of the algorithm, we have carried out simulation experiments from link bandwidth, mapping cost and virtual network request (VNR) acceptance rate. The final results show that the proposed algorithm is better than other representative algorithms in the above indicators.
\end{abstract}


\IEEEpeerreviewmaketitle

\section{Introduction}

With the continuous development and improvement of network technology and infrastructure, the communication mode between terminals gradually changes from wired communication to wireless network communication \cite{b1,zh1,b2,b3,n1,zh4,n2}. In the context of building a smart city, this transformation is particularly important. Wireless communication can overcome many inconveniences and defects of wired communication. For example, in the field of IoV, signal indication and resource allocation instructions are transmitted in the form of wireless signals. In the environment of IoV, the self driving vehicle needs to obtain road information and other environmental information independently. This requires all kinds of network infrastructure (satellite, server, signal station, etc.) to provide network resources for vehicles in time to ensure their normal operation  \cite{ch2,zh2,ch2a}. The transmission of these network resources needs a lot of bandwidth as a guarantee. Using wireless network communication to schedule network resources is the task of wireless communication \cite{ch4,ch2b}. Considering the limitation of traditional network architecture in resource scheduling, it is necessary to design a reliable scheme of resource scheduling using wireless communication mode in virtual network environment \cite{z4,d2,zh3}. Radio network resource management faces severe challenges, including storage, spectrum, computing resource allocation, and joint allocation of multiple resources \cite{jcx1,jcx2}. With the rapid development of communication networks, the integrated space-ground network has also become a key research object \cite{jcx3}.

In recent years, the emerging network virtualization (NV) can effectively make up for the shortcomings of existing networks and VNE is one of the key issues in NV research \cite{b4,c1}. Multi domain VNE is the focus of this field. According to the different needs of developers and the different goals of interests, the mapping goals will be different. There are mainly mappings with the minimum resource cost, the shortest time delay, the largest bandwidth, load balancing mapping and the largest revenue \cite{b5}.

\begin{figure}[!htbp]
\centering
\fbox{\includegraphics[width=0.88\columnwidth,height=.4\textwidth]{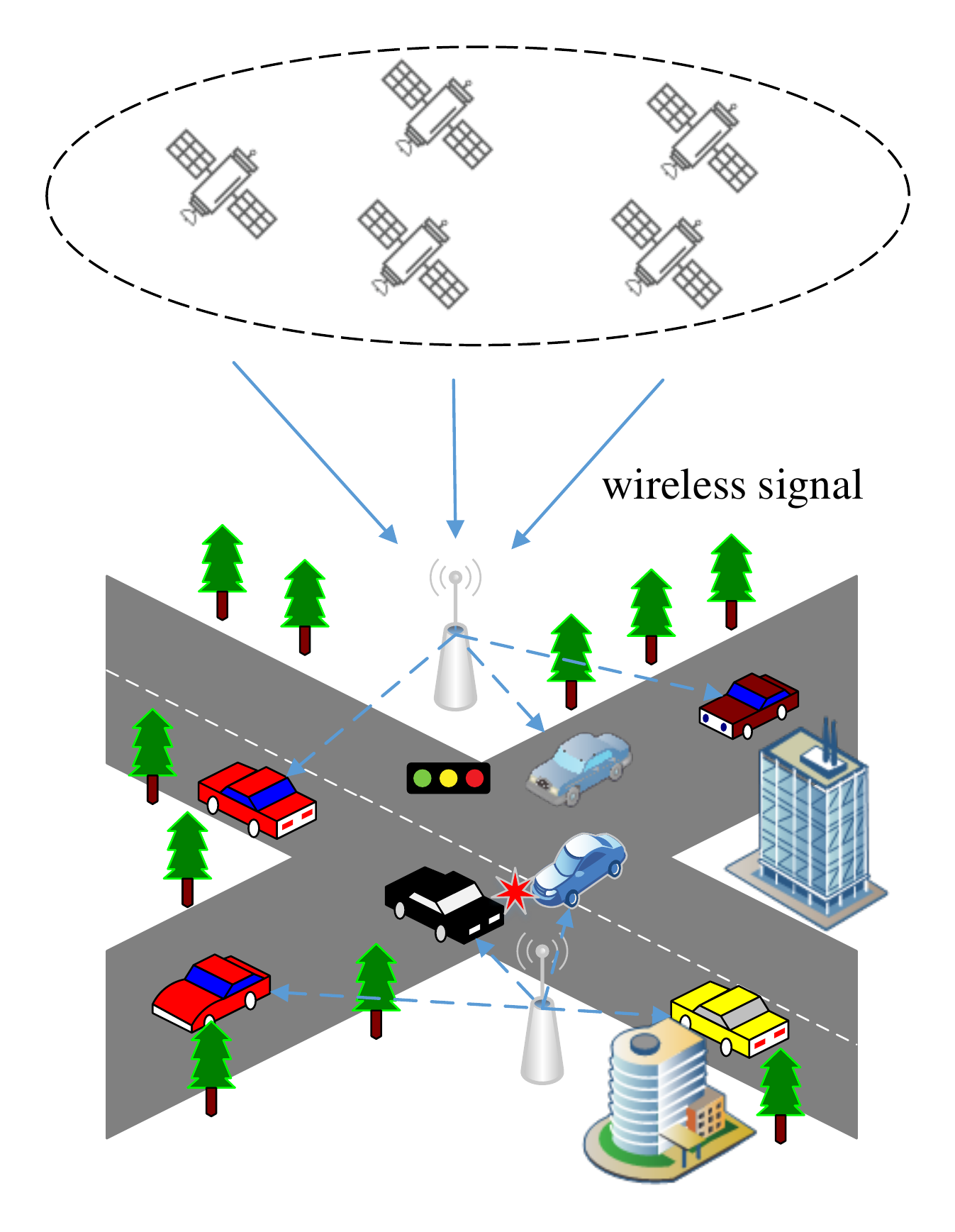}}
\caption{Wireless communication in IoV scenario.}
\label{fig_1}
\end{figure}

The application of IoV shows the necessity and importance of using wireless network communication to realize network resource scheduling. In IoV environment, the driver-less vehicles may require the support of a large number of network resources. Its application scenario is shown in the Fig. \ref{fig_1}. For example, the car needs to know the road conditions in advance, judge the distance from other vehicles and perceive the change of indicator light. The acquisition of these information happens over the network and the network bandwidth resources play a great role in the transmission of these information. In a short time, there may be a large number of driver-less vehicles on the road, who need to obtain road information simultaneously. This situation requires that the wireless network must be robust enough to transmit enough resource scheduling information. This kind of instantaneous and large amount of network resource demand will bring huge pressure to the underlying network infrastructure \cite{ch1,ch3}.

For IoV and other services with high bandwidth requirements, it is of great significance to study the bandwidth index in multi domain VNE. How to realize the resource scheduling of VNR in wireless communication environment is an urgent problem. With the blowout growth of network end-users, the network scale has increased dramatically. So the substrate network also becomes more and more complex. The VNE problem on a single domain cannot meet the needs of modern business more and more. A good solution to this problem is to transform the single-domain VNE problem into the multi-domain VNE problem. BA-VNE algorithm focus on bandwidth constraints and use PSO to optimize. Good results are achieved in the final experiment.

\subsection{Contributions}

The main contributions of this article are listed below.

(1) We propose a BA-VNE algorithm to solve the problem of insufficient bandwidth supply in the field of IoV. This algorithm focuses on the bandwidth index of multi domain VNE. Based on the objective function, the link with abundant bandwidth resources is selected for mapping.

(2) We adopt a centralized multi domain VNE architecture to improve the performance of the algorithm. We introduce the traditional natural heuristic algorithm PSO into BA-VNE algorithm. Variable factors are added to PSO to avoid the final mapping result falling into local optimum.

(3) We have done a mass of simulation experiments to verify the effectiveness of BA-VNE algorithm. BA-VNE algorithm is compared with other representative algorithms in the aspects of link bandwidth selection, mapping cost, VNR acceptance rate and link utilization. Experiments show that BA-VNE algorithm is effective.

\subsection{Organization}

The rest of the paper is organized as follows. Part 2 analyzes the research work of VNE problem. Part 3 introduces some related problems of multi-domain VNE. Part 4 introduces the model of BA-VNE algorithm and puts forward the evaluation index. Part 5 introduces the BA-VNE algorithm and gives the pseudo-code implementation. Part 6 introduces the setting of experimental environment and analyzes the experimental results. In the last part of the paper, we summarize the main content of the paper.

\section{Related Work}

\subsection{Centralized Multi-Domain VNE Algorithm}

There are different classification criteria for multi-domain VNE algorithms \cite{z3}. We classify multi-domain VNE algorithms according to centralized architecture and distributed architecture. The BA-VNE algorithm is based on the centralized network architecture. The embedding process of the algorithm is completed by the global controller and the local controller \cite{b6,z1,b7}.

Literature \cite{b8} proposed a multi-domain VNE scheme based on software-defined network (SDN). The mapping costs of nodes and links within the local controller are first estimated. Based on this cost, a candidate node selection algorithm is designed. Finally, the PSO algorithm is extended to the global controller to reduce the overall embedding cost. Therefore, the paper focuses on the influence of embedding cost on multi-domain VNE algorithm. Literature \cite{b9} proposed a cross-domain VNE algorithm based on minimum cost. The algorithm selects the virtual nodes to be mapped by pre-set constraints and selects the virtual links to be embedded by the minimum weight routing algorithm. Then the substrate path set with the least weight is selected by using Kruskal's minimum spanning tree and finally the embedding of virtual network is carried out. Therefore, the paper focuses on the influence of resource cost on the embedding algorithm of cross-domain virtual network. Literature \cite{b10} proposed a centralized cross-domain VNE algorithm. Based on dynamic topology perception and resource attributes, a network topology attribute evaluation model is established. The model is then used to measure the mapping priority of the nodes. Finally, the network load state is analyzed according to link resource cost. A meta heuristic algorithm based on iterative search is proposed in \cite{b11}. The algorithm mainly solves the complexity and scalability of multi-domain VNE. The main idea of the algorithm is to promote link embedding by node embedding. This can achieve cost effectiveness and help to decompose the network requests between cloud service providers in the network cloud environment online.

\subsection{Distributed Multi-Domain VNE Algorithm}

The distributed virtual network architecture is different from the centralized virtual network architecture. In the distributed VNE architecture, each InP is an independent individual and the information of each physical domain is not public. This is due to factors such as confidentiality and commercial competition among InPs. In this case, the VNE work needs the cooperation of InPs and service providers (SP) \cite{b12,z6,c3}.

In reference \cite{b13}, a mapping strategy based on bidding method is proposed. The strategy first assigns the virtual network embedding requests (VNER) from each SP to the InP. The main task of the InP is to select the subgraph of the VNER to be embedded, then send the rest of the VNER to other InPs. Reference \cite{b14} proposed the V-Mart architecture based on the authenticity of virtual resource auction. This architecture uses two-step Vickrey model to build a virtual environment similar to the bidding platform for InPs. Through bidding and negotiation among participating InPs, VNR decomposition is carried out to solve the problem of multi-domain VNE. Reference \cite{b15} proposed a policy-based VNE architecture VINEA. The algorithm separates the policy (target) from the underlying mapping mechanism (resource, virtual network mapping) and can contain existing embedding methods. It can be used to design VNE solutions for different scenarios just by instantiating different policies. The purpose of this approach is to improve the utilization of the underlying network resources, thereby maximizing InPs revenue. In addition, reference \cite{d1} also proposed a VNE algorithm based on PSO. The algorithm uses SDN architecture to solve the problem of VNE which focuses on embedding cost. So to sum up, the existing multi-domain VNE algorithms, whether centralized or distributed, mainly focus on embedding cost, embedding cost, embedding success rate and other indicators. So far as we know, no multi-domain VNE algorithm has focused on bandwidth as a constraint.

\section{Problem Specification}

\subsection{Description of the Basic Problem of Multi-Domain VNE}

The problem of VNE can be divided into two parts: embedding in a single network domain and embedding across multiple underlying networks. The latter is more complex than the former. The complexity is reflected in the fact that the entire underlying infrastructure passes through multiple management domains or data centers. Multiple InPs work together to form the underlying infrastructure network, while different InPs manage their respective underlying nodes and links. Information between InPs is not disclosed to each other. Therefore, the pricing of link resources between multiple InPs is not uniform, which makes it more difficult to determine the bandwidth.

In the cross domain VNE architecture, the SP will send users' VNRs to the underlying network to request network resources. The global controller will segment the VNRs and distribute them to the local controller. Finally, the local controller will allocate the network resources needed for VNE to each network request. Given a VNR and the underlying infrastructure composed of multiple InP management domains, the embedding of the virtual network to the underlying infrastructure is accomplished with the goal of seeking the maximum bandwidth resource link. The InP does not publicly disclose link resource information within its managed domain, nor does it determine resource information on inter-domain links. Therefore, the decision maker of the VNE does not have global information about the entire underlying infrastructure, which makes it difficult to determine the maximum bandwidth.

The underlying layer of the multi-domain VNE framework is composed of multiple substrate networks. A VNE request may be mapped to multiple substrate networks. Each underlying network is managed by a corresponding InP. Each underlying network has logic controllers (centralized or distributed) to perform management and control functions for infrastructure vendors. Each boundary node in the different underlying network is connected by inter-domain link to form the substrate network.

There is a limit to the amount of information that can be uploaded from the local controller to the global control. If all the node and link information of the underlying network is uploaded to the global controller, it will inevitably cause a lot of overhead. Therefore, only the following three kinds of information are uploaded to the global controller:

(1) Available resources and unit price for candidate nodes (This part is described in part 3.2).

(2) Available bandwidth resources and unit price of links between physical domains.

(3) Link bandwidth resource aggregation unit from candidate node to boundary node in the domain. That is, only boundary nodes are kept in a physical domain and the remaining nodes are replaced by a non-boundary node. All links are also aggregated between the boundary node and the non-boundary node. The in-domain link unit price is the link aggregation unit price.

\begin{figure}[!htbp]
\fbox{\includegraphics[width=0.95\columnwidth]{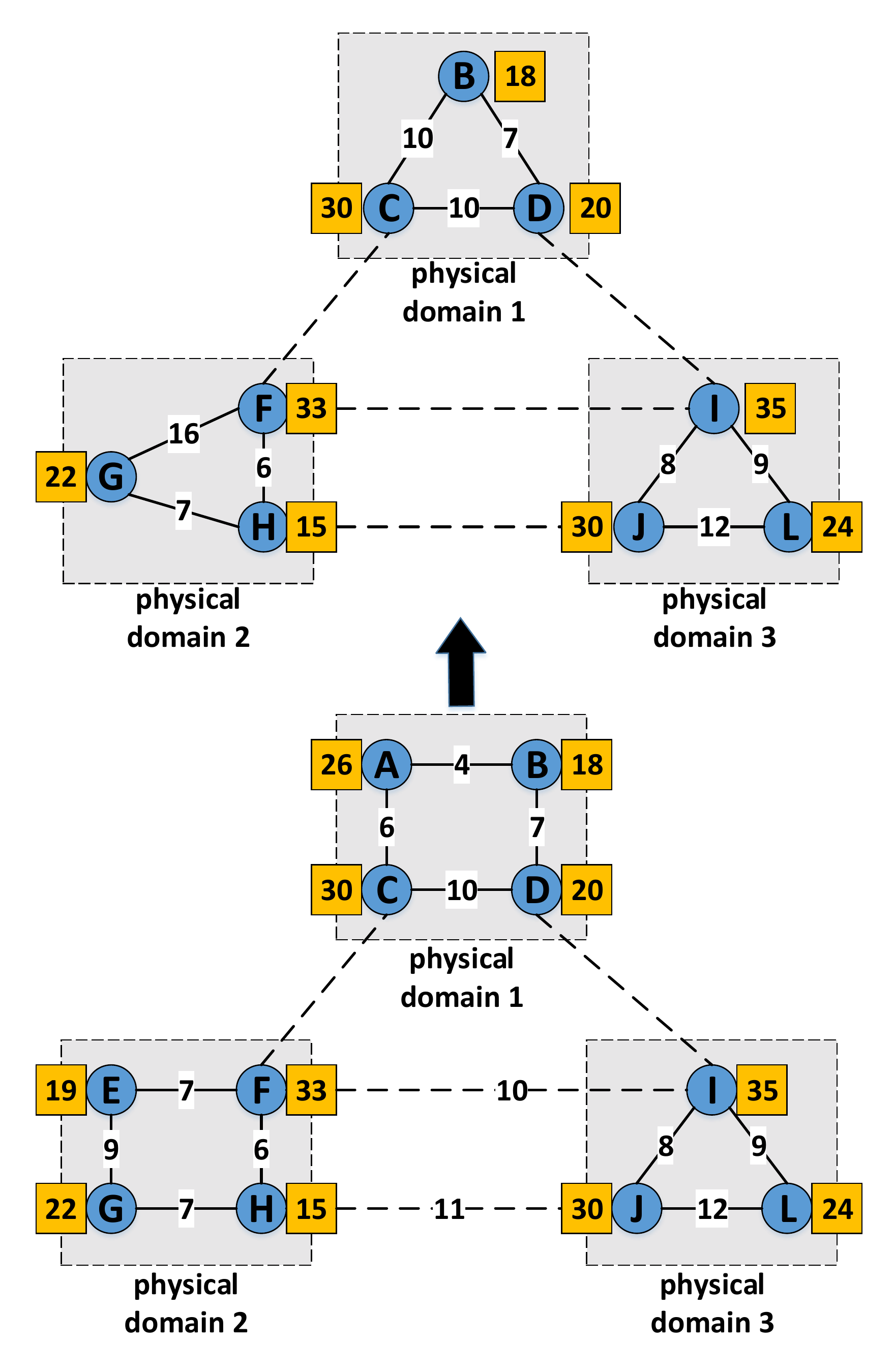}}
\caption{local controller uploads information}
\label{fig_2}
\end{figure}

Taking Fig. \ref{fig_2} as an example, link aggregation occurs in both physical domain 1 and physical domain 2 during information upload. For physical domain 1, since only boundary nodes and one non boundary nodes are reserved, only nodes \textit{B}, \textit{C} and \textit{D} are uploaded to the global controller. Before uploading the information, there is a link connection between node \textit{A}, node \textit{B} and node \textit{C}. The upper bandwidth are 4 and 6. Because of the link aggregation, the two links originally connected to node \textit{A} are aggregated into one link. The bandwidth value of the link also changes from the sum of the two aggregated link bandwidth values to 10. Physical domain 2 also has link aggregation when uploading information. Only boundary nodes \textit{F} and \textit{H} and non boundary nodes \textit{G} are reserved.

\subsection{Selection of Candidate Nodes}

The specific steps for selecting candidate nodes are as follows:

(1) Select the boundary nodes in the candidate domain as the candidate nodes until all boundary nodes are selected.

(2) Select the non boundary nodes with substrate link with the boundary nodes as the candidate nodes. The substrate link between the two nodes is saved and the bandwidth value on the link is recorded at the same time.

(3) If there is no link between the non boundary node and the boundary node, continue to select the next hop node of the boundary node.

Repeat the above steps until all nodes in the candidate domain are selected. Finally, the bandwidth values of all links in the domain are recorded.

\begin{equation}
\begin{aligned}
bandwidth(n_{border}^s,n_i^s) \neq 0.
\end{aligned}
\end{equation}

The selection condition is shown in Eq. (1). As long as the link bandwidth value between the boundary node and a node is not 0, there is a substrate link between the two nodes. Then both nodes can be used as candidate nodes and the substrate link between them can be saved.

\section{Network Models and Evaluation Indicators}

\subsection{Underlying Network Model}

The underlying network model can be represented by $G^s=\{N^s,L^s\}$, specifically, this is a weighted undirected graph. $G^s$ represents the entire underlying network. $N^s$ represents all the underlying network nodes and $L^s$ represents all the underlying network links. The computing power of the substrate node $n^s \in N^s$ is expressed in terms of $CPU(n^s)$. The substrate link between the substrate nodes $x$ and $y$ is represented by $l^s(x,y)$, where $l^s(x,y) \in L^s$. The bandwidth capacity of the substrate link is expressed in terms of $BW(l^s)$.

\begin{figure}[!htbp]
\fbox{\includegraphics[width=0.95\columnwidth]{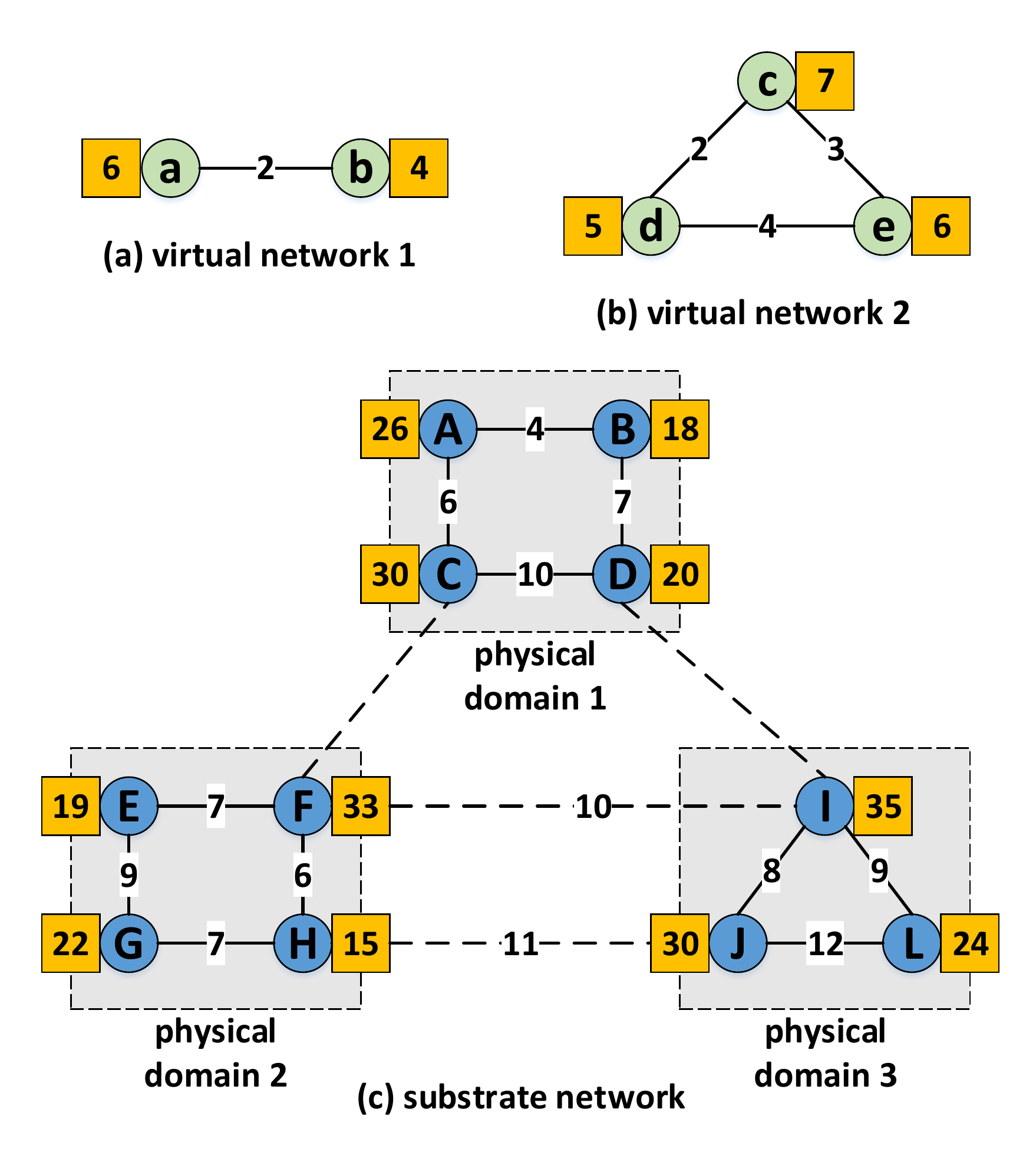}}
\caption{diagram of virtual network and substrate network}
\label{fig_3}
\end{figure}

Fig. \ref{fig_3} (c) is an example diagram of a substrate network. The circles represent the substrate nodes. The amount of computational resources available for the substrate node is represented by the number in the rectangle. The connections between the nodes represent substrate links. The amount of bandwidth available for the substrate link is represented by the number of connections between the substrate nodes.

\subsection{Virtual Network Model}

The virtual network model can be represented by $G^v=\{N^v,L^v\}$ which is also a weighted undirected graph. Where $G^v$ represents the entire virtual network, $N^v$ represents all the virtual network nodes and $L^v$ represents all the virtual network links. The computing resource requirements of virtual node $n^v \in N^v$ is expressed in terms of $CPU(n^v)$. The virtual link between the virtual nodes $x$ and $y$ is represented by $l^v(x,y)$, where $l^v(x,y) \in L^v$ has a bandwidth requirement of $BW(l^v)$.

Both (a) and (b) in Fig. \ref{fig_3} are examples of virtual networks. The circle represents the virtual node. The computing resource requirements of virtual nodes are represented by numbers in the rectangle. The lines between the nodes represent virtual links. The bandwidth resource demand of virtual link is represented by the number of the connection between virtual nodes.

\subsection{VNR Model}

The VNR first arrives at the global controller and subnet partition is performed under the operation of the global controller. A virtual node is divided into a separate subgraph of cells. The VNR can be represented as $G^v=\{N^v,L^v,C_N^v,C_L^v,D_N^v\}$. Where $G^v$ represents the entire VNR process. $N^v$ represents all the virtual network nodes and $L^v$ represents all the virtual network links. $C_N^v$ represents the computational footprint of the virtual node, which is the CPU footprint. $C_L^v$ represents the bandwidth resource demand of the virtual link. $D_N^v$ represents the candidate domain of the virtual node.

\begin{figure}[!htbp]
\fbox{\includegraphics[width=0.97\columnwidth]{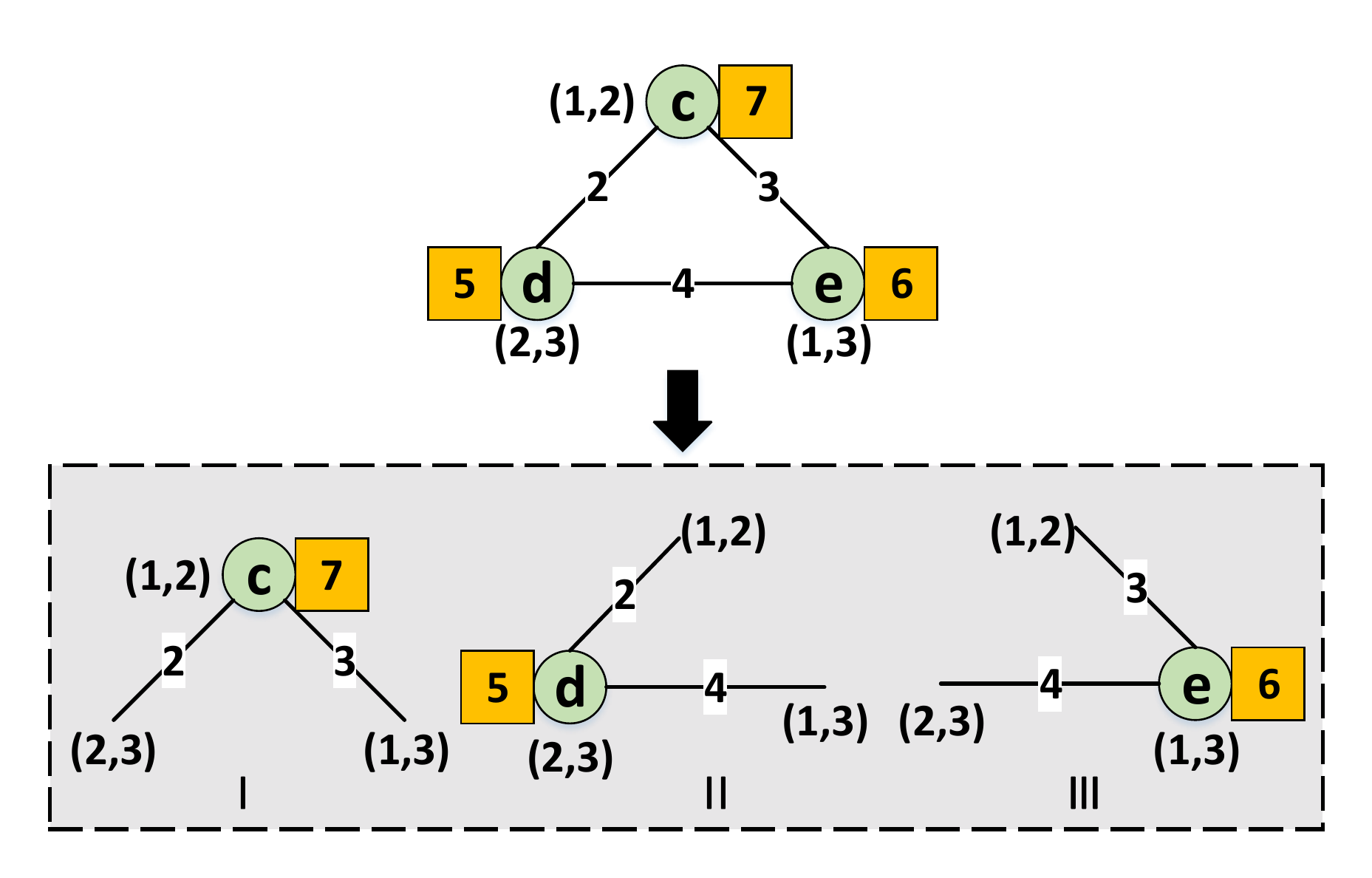}}
\caption{virtual network requests and their subgraphs}
\label{fig_4}
\end{figure}

VNRs and their subgraphs are shown in Fig. \ref{fig_4}. The virtual network with three virtual nodes and three virtual links is divided into three VNR subgraphs. The subgraph includes information such as the computation resource demand, the candidate domain number of virtual node and the bandwidth demand of virtual link.

Take the example of the VNR subgraph \uppercase\expandafter{\romannumeral1}. The candidate domains of virtual node \textit{c} include physical domain 1 and physical domain 2 which means that virtual node \textit{c} can only select these two physical domains as candidate domains. The two virtual links connected to the node correspond to the candidate domains of virtual node \textit{d} and virtual node \textit{e} respectively. The other two virtual nodes are similar.

\subsection{The Objective Function}

In order to ensure that all virtual nodes can be embedded, we need to select the substrate links with the largest remaining bandwidth resources to map to meet the service requirements of the IoV. Here, we define a mapping target formula:

\begin{equation}
\begin{aligned}
BW[l^s(u,v)] \geq \frac{\sum\limits_{L^s}BW(l_i^s)}{count+1}.
\end{aligned}
\end{equation}

We initialize the substrate network before selecting the candidate nodes. Select the substrate links that meet the standards and the selection criteria are shown in Eq. (2). First, sum up the remaining bandwidth resources of all links in a physical domain to obtain the total bandwidth resources of the links in the domain. Then the total link bandwidth resource is divided by the number of links in the domain and the mean value is taken as a selection criterion. It can be considered that if the bandwidth value of a link is greater than the average value, then the link is eligible. Where, the denominator $count$ represents the number of substrate links, plus 1 is to prevent the situation that the denominator is 0. The numerator is the sum of the bandwidth of all links in the domain. On the left side of the formula is the bandwidth value on the link between the substrate nodes $u$ and $v$.

In Fig. \ref{fig_4} of the VNR subgraph \uppercase\expandafter{\romannumeral1} and Fig. \ref{fig_2} by the local controller in the upload of the substrate network as an example, shows the specific meaning of the Eq. (2). Suppose virtual node \textit{c} selects substrate node \textit{A} in candidate domain 1 for mapping. Since substrate node \textit{A} is not uploaded, it should be calculated through substrate node \textit{B} at this time. First, the bandwidth mean value $\frac{10+10+7}{3}$ in candidate domain 1 is calculated, then the mean value is compared with the value of all link bandwidth in that domain. The bandwidth value of the link between substrate node \textit{B} and \textit{C} is 10. The bandwidth value of the link between substrate node \textit{C} and \textit{D} is also 10. The bandwidth value of the two links is larger than the average value, so they can be used as candidate links. However, the bandwidth value of the link between substrate node \textit{B} and \textit{D} is 7, which is less than the average value, so this link cannot be used as a candidate link. In this form, the local controller passes all substrate nodes and links to boundary nodes that meet the bandwidth value requirements up to the global controller. This completes the selection of candidate nodes. Eq. (2) shall meet the following conditions:

\begin{eqnarray}
max[CPU(n^s) \geq CPU(n^v)],\\
max[BW(l^s) \geq BW(l^v)],\\
num(n_v^s) = 1,\\
D(n^s) = D(n^v),\\
n_{v1}^s = n_{v_2}^s \quad if \quad v_1 = v_2,\\
p(n_i^s,n_k^s) = p(n_k^s,n_i^s).
\end{eqnarray}

Among them, Eq. (3) indicates that the computing resource of substrate node $n^s$ should not be less than that of virtual node $n^v$. Eq. (4) indicates that the bandwidth resource of substrate link $l^s$ should not be less than that of virtual link $l^v$. Eq. (5) indicates that a virtual node $n^v$ in a VNR can only be embedded in one substrate node $n^s$. Eq. (6) indicates that the target substrate node of the virtual node must be within the candidate domain. Eq. (7) indicates that a substrate node can only be mapped by one virtual node. Eq. (8) indicates that the path of the substrate link has no directionality.

\subsection{The Evaluation Index}

In this part, we formulate the evaluation index of the VNE algorithm.

InPs charge according to the amount of underlying resources consumed by users. In VNE problem, CPU resources and bandwidth resources are the main resources consumed. Therefore, the cost of VNE is represented by the consumption of CPU resources and bandwidth resources. The comprehensive mapping cost is expressed as follows:

\begin{equation}
\begin{aligned}
Cost=\sum_{num(n^v)}CPU(n^v) \times Cost(n_v^s)+ \\ \sum_{num(l^v)} BW(L^v) \times Cost(l_v^s),
\end{aligned}
\end{equation}
where, $CPU(n^v)$ represents the computing resource demand of virtual node $n^v$. $BW(L^v)$ represents the bandwidth resource demand of virtual link $l^v$. $n_v^s$ represents that the substrate node of the final mapping of virtual node $n^v$ is $n^s$. $l_v^s$ represents that virtual link $l^v$ is mapped to substrate link set $l^s$. Because a virtual link can be split and mapped to different substrate links. $Cost(n_v^s)$ represents the computing resource unit price of substrate node $n^s$. $Cost(l_v^s)$ represents the bandwidth resource unit price of substrate path $l^s$.

The VNE request acceptance rate is shown as follows:

\begin{equation}
\begin{aligned}
Acp=\lim_{x \to \infty} \frac{\sum_{n=1}^{N}Acp(G^V,n)}{\sum_{n=1}^{N}All(G^V,n)}.
\end{aligned}
\end{equation}

The acceptance rate of VNRs refers to the proportion of successful mapping of VNRs. Where $n$ represents the number of virtual networks. The ratio of the number of successful virtual networks $\sum_{n=1}^{N}Acp(G^V,n)$ to the total number of VNRs that arrive $\sum_{n=1}^{N}All(G^V,n)$.

Link utilization is shown as follows:

\begin{equation}
\begin{aligned}
LinkUse=\lim_{x \to \infty} \frac{\sum_{n=1}^{N}Num(L^V \to L^S)}{All(L^S)}.
\end{aligned}
\end{equation}

Link utilization refers to the proportion of mapped substrate links to all substrate links. The ratio of the number of successful virtual links $\sum_{n=1}^{N}Num(L^V \to L^S)$ to the total number of links $All(L^S)$ in the substrate network.

\section{Algorithm Description and Implementation}

\subsection{Algorithm Description}

We divide the implementation of BA-VNE algorithm into the following steps.

(1) The specific steps of candidate node selection are described in section 3.2. We take the pre mapping cost of virtual nodes as the selection criteria of candidate nodes. The pre mapping cost of node can be expressed as follows:
\begin{equation}
\begin{aligned}
Cost_{forecast}=CPU(n^v) \times Cost(n_v^s).
\end{aligned}
\end{equation}

(2) VNR pre-mapping: The local controller sends the selected set of candidate nodes to the global controller. The global controller pre maps the VNR based on this information. If the premapping is successful, the premapping results are delivered to the local controller.

(3) Intra-domain embedding of VNRs: The local controller receives the pre-mapping result from the global controller, then carries out intra-domain node mapping. This process includes the embedding of virtual nodes and links. After the virtual node is successfully embedded, the intra-domain link with the most remaining bandwidth resources around the node is selected for embedding.

(4) Inter-domain link embedding: If all intra-domain embeds are successful, the global controller completes the inter-domain link embedding. The basis is to first select the substrate link with the most remaining bandwidth resources for embedding. Finally, the embedding process is finished.

\begin{figure}[!htbp]
\centering
\includegraphics[width=1.0\columnwidth,height=.75\textwidth]{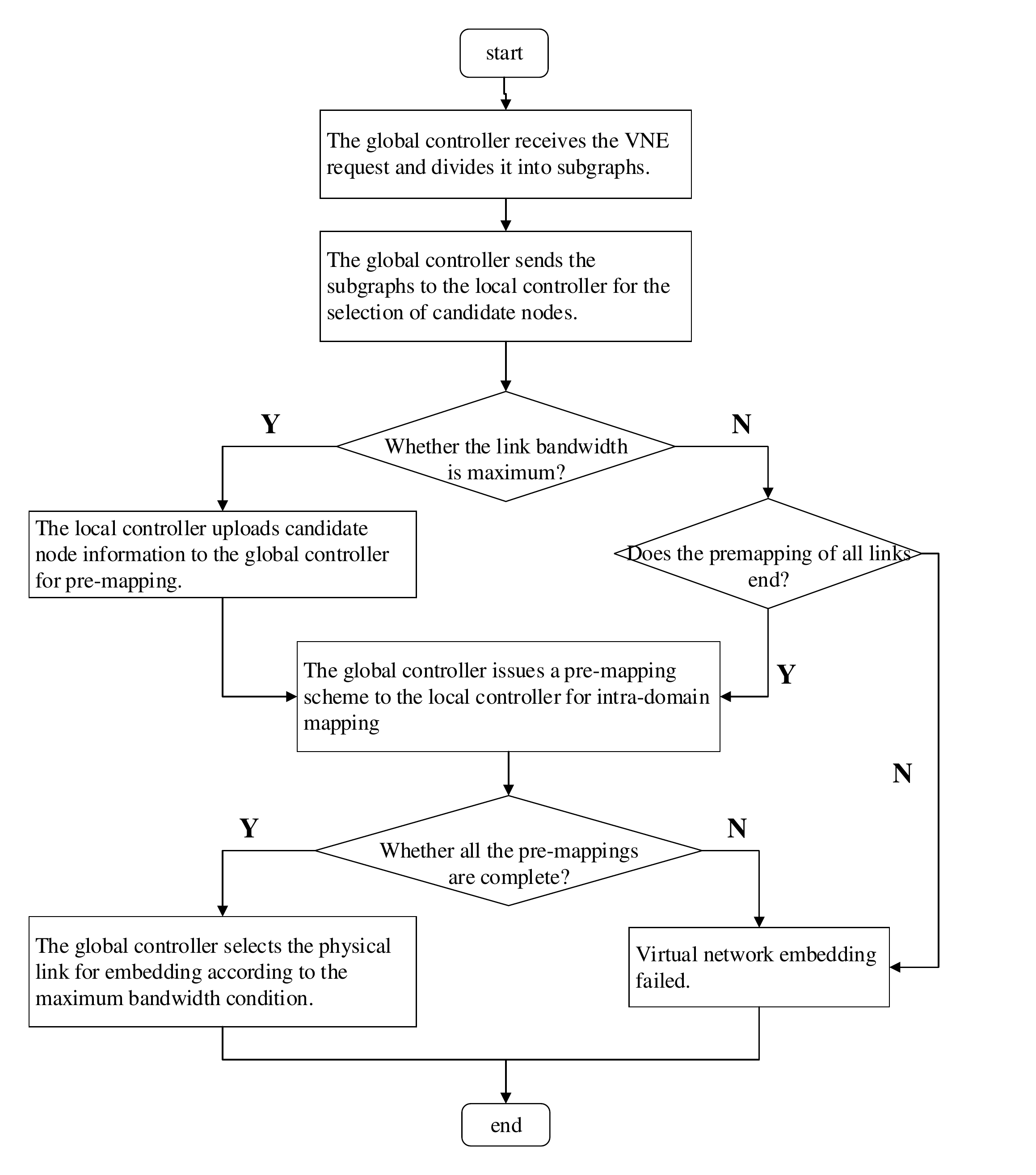}
\caption{BA-VNE algorithm flowchart}
\label{fig_5}
\end{figure}

The flowchart of BA-VNE algorithm is shown in Fig. \ref{fig_5}.

\subsection{Algorithm Implementation}

In this part, we will describe the implementation process of two important parts of BA-VNE algorithm in detail. These two steps are candidate nodes selection and VNR pre mapping. We give their pseudo code implementation.

\textbf{1. Candidate node selection algorithm}: The second line of the algorithm represents setting the substrate nodes to a state where none of them are mapped, which is initialization. Because one substrate node can only host one virtual node from the same VNR, if the substrate node has been mapped by a virtual node in a VNR, initialization must be carried out to prevent affecting the node mapping in other VNRs. Line 4 represents the node loop operation (if there are boundary nodes, judge them first). Line 5 first determines whether a link exists between two nodes. If so, line 6 determines whether the remaining bandwidth resource found between two nodes is greater than the average link bandwidth of the physical domain. Because the problem to be studied is multi-domain VNE, inter-domain link embedding is necessary, so the boundary nodes must be embedded. Line 7 indicates that the selected candidate node is saved.
\begin{algorithm}[!h]
	\caption{Candidate Node Selection Algorithm}
	\begin{algorithmic}[1]
		\Require
		{$G_i^s=\{N_i^s,L_i^s\}$,$G_i^s=\{N_i^v,L_i^v\}$};
		\Ensure
		{$N_{i,candinode}^s$};
		\Procedure{}{}
		\State $\phi(N^s)\leftarrow0$;
		\EndProcedure
		\For {$all \quad n_i^s \in N_i^s$}
		\If {$bandwidth{n_{border}^s,n_i^s}\neq0$}
		\If {$BW[l^s(u,v)] \geq ave\_bandwidth$}
		\State $n_{candinode}^s = (n_{border}^s,n_i^s)$;
		\EndIf
		\EndIf
		\EndFor
		\State \Return $n_{candinode}^s$;
	\end{algorithmic}
\end{algorithm}

\textbf{2. VNR pre-mapping}: After the global controller receives the candidate node information (including the case where it is connected to the boundary node), this information is combined with inter-domain links to form a new network topology called global candidate node network. Based on this network, PSO is used to realize the premapping of multi-domain VNE scheme with maximum bandwidth. PSO is a typical natural heuristic algorithm, which is based on the analogy of the foraging process of birds. It is often used to solve global search problems to get the optimal solution \cite{b16,b17,z2}. In the discrete PSO algorithm, the calculation formula of particle position and velocity is as follows:
\begin{equation}
\begin{aligned}
v_i^{new}=v_i+c_1r_1(x_i^{pb}-x_i)+c_2r_2(x^{gb}-x_i),
\end{aligned}
\end{equation}

\begin{equation}
\begin{aligned}
x_i^{new}=x_i+v_i^{new}.
\end{aligned}
\end{equation}

There is a concept of fitness function in PSO, which we express as Eq. (2). $v_i$ originally represents the speed of particles in PSO. In BA-VNE algorithm, $v_i$ represents the change direction of the ith mapping scheme. $c_1$ and $c_2$ represent the learning factors in PSO algorithm. Reasonable setting of the learning factors can improve the efficiency of the algorithm. $r_1$ and $r_2$ represent random numbers between 0 and 1, which are variable factors. $x_i^{pb}$ is meant to represent the best position of a particle in PSO. In BA-VNE algorithm, the fitness function representing this mapping scheme is optimal. $x^{gb}$ is meant to represent the global optimal position in PSO. In BA-VNE algorithm, the fitness function of this mapping scheme is the best among all particle mapping schemes.

\textbf{The steps of PSO are as follows:}

(1) \textit{Initialize}: Location and velocity of randomly generated particles to produce an embedding scheme.

(2) \textit{Evaluate each particle}: Calculate the fitness value of the particle and the embedding cost of the embedding scheme.

(3) \textit{Update}: Velocity \& position of particle --$>$ Eqs. (13-14).

(4) \textit{Check whether the end conditions are met.} If the maximum number of iterations has been reached, PSO algorithm will stop and output the optimal solution. Otherwise, go back to step (2) and continue.

\begin{algorithm}[!h]
	\caption{Virtual Network Request Premapping}
	\begin{algorithmic}[1]
		\Require
		{$G^v=\{N^v,L^v\}$,$PseudoTopo$};
		\Ensure
		{$premapping(n^v)$};
		\Procedure{}{}
		\State $\phi(N^s)\leftarrow0$;
		\EndProcedure
		\For {$i < Max \quad Iteration$}
		\For {$j < Particial \quad Number$}
		\State $X_{i+1} = Update(X_i)$;
		\State $V_{i+1} = Update(V_i)$;
		\If {$Cost(X_{i+1}) < Cost(X^{pb})$}
		\State {$X^{pb} = X_{i+1}$};
		\EndIf
		\If {$Cost(X^{pb}) < Cost(X^{gb})$}
		\State {$X^{gb} = X^{pb}$};
		\EndIf
		\EndFor
		\EndFor
		\State \Return $premapping(n^v)$;
	\end{algorithmic}
\end{algorithm}

The second line of the algorithm represents setting the candidate substrate node to a state that is not embedded, that is, initialization. Line 4 represents looping through the maximum number of iterations. Line 5 is a loop for the number of particles. Line 6 updates the position of the particle. Line 7 is the update particle velocity. Lines 8 to 12 are the judgment and update operation of particle position.

\subsection{Algorithm Complexity}

The complexity of BA-VNE algorithm is mainly caused by candidate node selection process, PSO algorithm optimization process and mapping process, so its overall complexity is $O(n \times (2 \times |N^s| + |L^s|) + (|N^v|+|L^v|))$. Where $n$ represents the number of physical domains. $|N^s|$ represents the number of nodes in different physical domains. $|L^s|$ is the number of links in different physical domains. $|N^v|$ is the number of all virtual nodes. $|L^v|$ is the number of all virtual links.

\section{Performance Evaluation}

This chapter first introduces the environment of simulation experiment and the parameter setting of simulation experiment. Then the experimental results are presented and analyzed. We compare the BA-VNE algorithm proposed in this paper with several representative algorithms proposed in literature \cite{b8}, literature \cite{b9}, literature \cite{b18} and literature \cite{b19} to verify the performance of BA-VNE algorithm. The main ideas are listed in TABLE \uppercase\expandafter{\romannumeral1}. Specifically, we compare bandwidth selection, mapping cost and VNR acceptance rate.

\begin{table}
\caption{Algorithm ideas}
\renewcommand\arraystretch{1.5}
\begin{tabular}{|p{15mm}|p{65mm}|}
\hline
Notation & The algorithm description\\
\hline
BA-VNE & The candidate nodes are selected according to the maximum bandwidth. PSO with genetic variation factors was used to solve the node embedding scheme. \\
\hline
VNE-PSO & The hop number from the boundary node is used as the selection criterion of candidate nodes. PSO algorithm is used to optimize the VNE scheme and the best VNE scheme is used as the final algorithm. \\
\hline
MC-VNM & Based on the minimum mapping cost, the virtual link is mapped first, then the virtual node is mapped. \\
\hline
LID-VNE & The bandwidth requirement is taken as the VNE index. Each network domain performs node mapping and link mapping according to different bandwidth requirements. \\
\hline
MP-VNE & A multi-objective VNE scheme transforms the multi-objective problem into a single objective problem. PSO algorithm is used to optimize the mapping scheme. \\
\hline
\end{tabular}
\end{table}

\subsection{Experiment Environment and Parameter Setting}

The simulation experiment was done on a computer with 8GB memory, 64-bit width and win8 operating system. We use traditional GT-ITM tools to generate the required network topology. The analysis of the experimental results and the drawing of the line chart were completed by Origin 8.5. The important parameters in the network topology are set as follows:\\
\indent In order to compare the experimental results in different numbers of data fields, we set up a total of four physical fields. The number of nodes in each physical domain is 30 and there are 2 boundary nodes in each physical domain. The amount of resources available for each substrate node is uniformly distributed [100,300]. The unit price of substrate node resources obeys the uniform distribution of [1,10]. The initial value of available bandwidth resources of the substrate link is set between 1000 and 3000. Link bandwidth resource unit prices follow uniform distribution [1,10]. The substrate link delay obeys the uniform distribution of [1,10]. The unit prices of inter-domain link bandwidth resources follow uniform distribution [5,15]. Inter-domain link delay obeys uniform distribution [10,30]. Substrate links are connected between substrate nodes with a 50\% probability.\\
\indent The candidate fields for each virtual node are set to 2. CPU demand for virtual nodes is evenly distributed [1,10]. The bandwidth resource demand of the virtual link is uniformly distributed [1,10]. The node connection probability in the VNR is set to 50\%. The arrival process of VNRs simulates Poisson distribution \cite{b20}. The summary of simulation parameters is shown in TABLE \uppercase\expandafter{\romannumeral2}.

\begin{table}[!h]
\caption{Summary of parameter settings}
\renewcommand\arraystretch{1.5}
\begin{tabular}{|p{60mm}|p{20mm}|}
\hline
Attribute name & Attribute value \\
\hline
number of physical domains & 4 \\
\hline
number of nodes per physical domain & 30 \\
\hline
computing resources of substrate nodes & U[100, 300] \\
\hline
computing resource unit price of substrate node & U[1, 10] \\
\hline
bandwidth resources of substrate link & U[1000, 3000] \\
\hline
bandwidth resource unit price of substrate link & U[1,10] \\
\hline
delay of substrate link & U[1,10] \\
\hline
number of boundary nodes in each physical domain & 2 \\
\hline
connection rate between substrate nodes & 50\% \\
\hline
price of inter-domain link bandwidth resource & U[5, 15] \\
\hline
interdomain link delay & U[10, 30] \\
\hline
number of candidate fields for each virtual node & 2 \\
\hline
computing resource requirements of virtual nodes & U[1,10] \\
\hline
bandwidth resource requirements of virtual link & U[1,10] \\
\hline
connection rate between virtual nodes & 50\% \\
\hline
\end{tabular}
\end{table}

\subsection{Experimental Results and Analysis}

We designed six comparative experiments to evaluate the performance of BA-VNE algorithm. The first two experiments are to prove the effectiveness of BA-VNE algorithm in selecting the largest possible bandwidth link. The following four experiments prove that the BA-VNE algorithm can guarantee the maximum bandwidth in terms of the synthetic mapping cost, VNR acceptance rate, average mapping delay and link utilization. The performance in these indicators is also excellent.

\textbf{Experiment 1:} Comparison of the average bandwidth of the selected substrate link at BA-VNE with the average bandwidth in the physical domain.

\begin{figure}[!h]
\includegraphics[width=1.0\columnwidth]{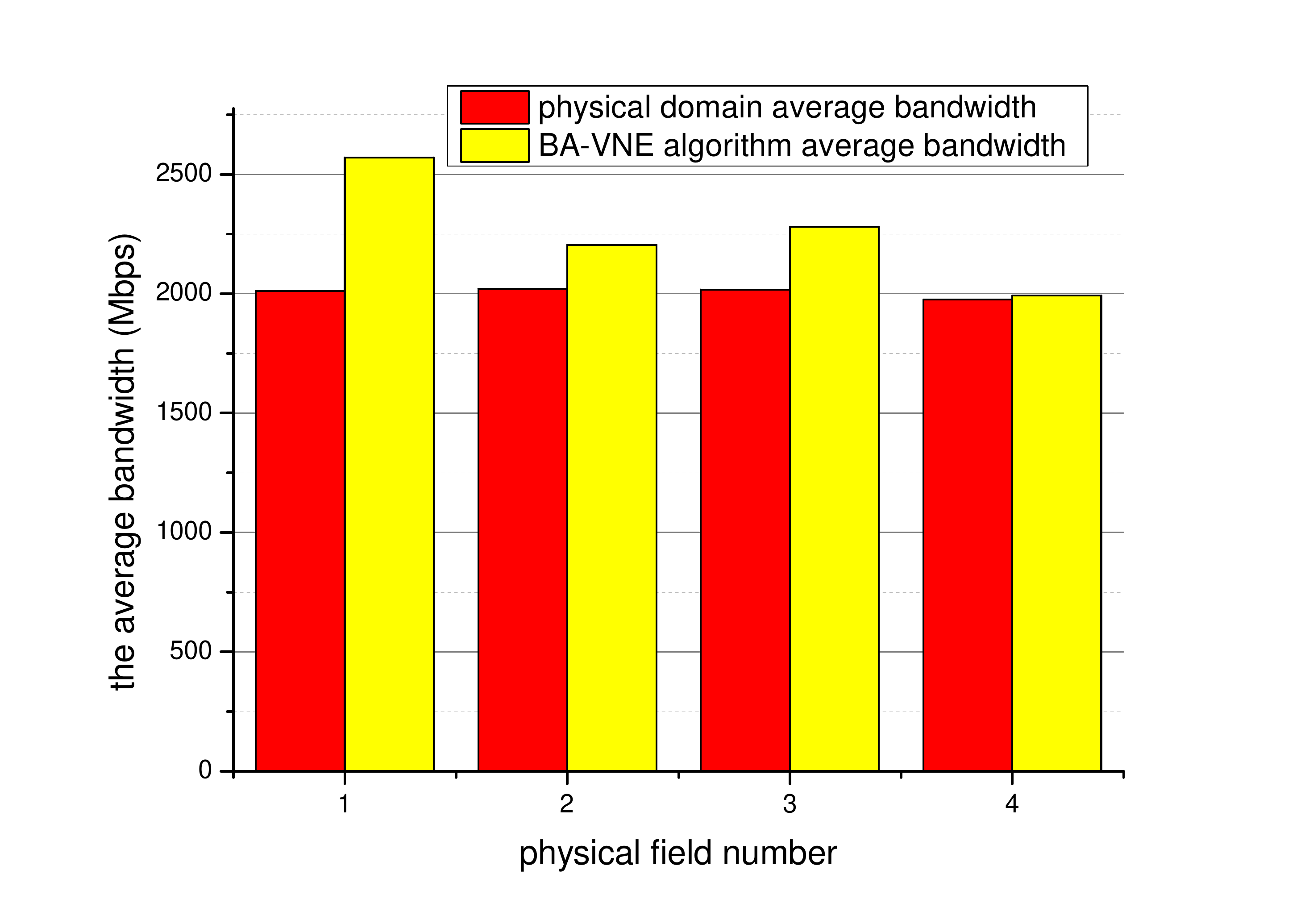}
\caption{The average bandwidth}
\label{fig_6}
\end{figure}

In this experiment, the average bandwidth of links in each physical domain is compared with the average bandwidth of links in each domain obtained by the BA-VNE algorithm. The average bandwidth of all links is obtained from the right side of Eq. (2). The average bandwidth of BA-VNE algorithm is measured experimentally. The experimental results are shown in Fig. \ref{fig_6}. The BA-VNE algorithm is effective and the average selected link bandwidth is significantly higher than the average bandwidth in each domain.

\textit{Analysis of experimental results:} The average bandwidth of the link selected by the BA-VNE algorithm is significantly higher than the average bandwidth of each domain. After calculation, the average in-domain link bandwidth from physical domain 1 to physical domain 4 is 2012, 2021, 2018 and 1976, respectively. According to Eq. (2), all substrate links with bandwidth resource less than this value cannot be selected as physical links. Therefore, in the results of each virtual network mapping request, the average link bandwidth in the four physical domains is larger than the above values by using the BA-VNE algorithm.

\textbf{Experiment 2: }Compare the average bandwidth with MP-VNE algorithm.

\begin{figure}[!h]
\includegraphics[width=1.0\columnwidth]{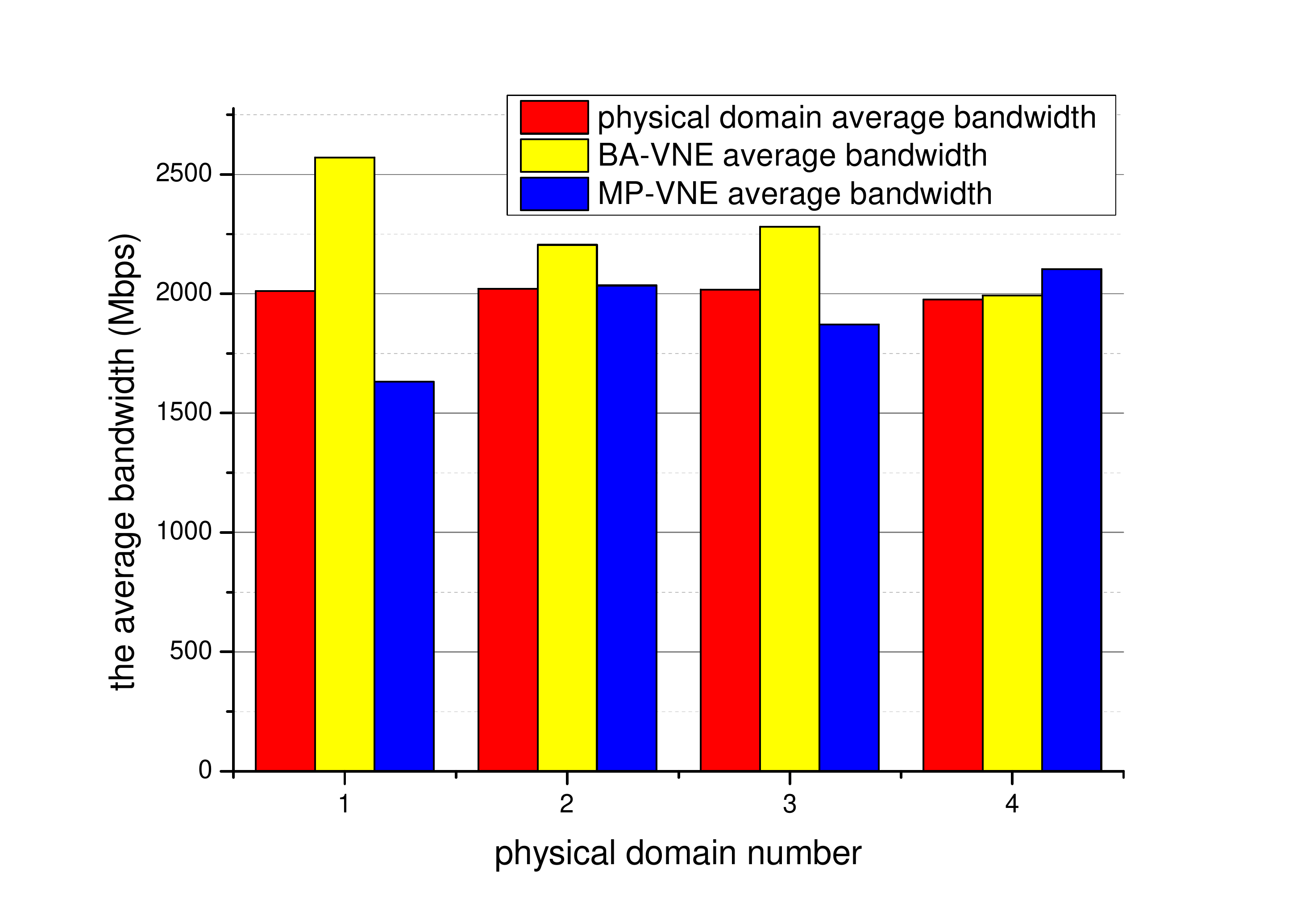}
\caption{Mean bandwidth value}
\label{fig_7}
\end{figure}

The purpose of this experiment is to show that BA-VNE algorithm has obvious advantage over other algorithms in choosing substrate links with large bandwidth for embedding. The average bandwidth of the two algorithms was compared in four physical domains. The comparison result is shown in Fig. \ref{fig_7}. The average link bandwidth selected by BA-VNE algorithm is significantly higher than the average link bandwidth selected by MP-VNE algorithm, while the MP-VNE algorithm has randomness in the selection of embedded link bandwidth.

\textit{Analysis of experimental results:} The BA-VNE algorithm focuses on the index of maximum link bandwidth. With Eq. (2) as the selection criterion, only substrate links whose bandwidth value is greater than the mean value of substrate links can be embedded. Therefore, the value of substrate link bandwidth selected by BA-VNE algorithm is generally the largest. The MP-VNE algorithm uses weights to transform the multi-objective optimization problem into a single-objective optimization problem. Select the boundary node using the estimated mapping cost formula. So it does not focus on bandwidth maximization. Therefore, the experimental results of MP-VNE algorithm are random when the average bandwidth is taken as the evaluation standard. It is not guaranteed that the link with the maximum bandwidth will be selected for embedding in every VNE process.

\textbf{Experiment 3:} Comparison of algorithm synthetic embedding cost.

\begin{figure}[!h]
	\includegraphics[width=1.0\columnwidth]{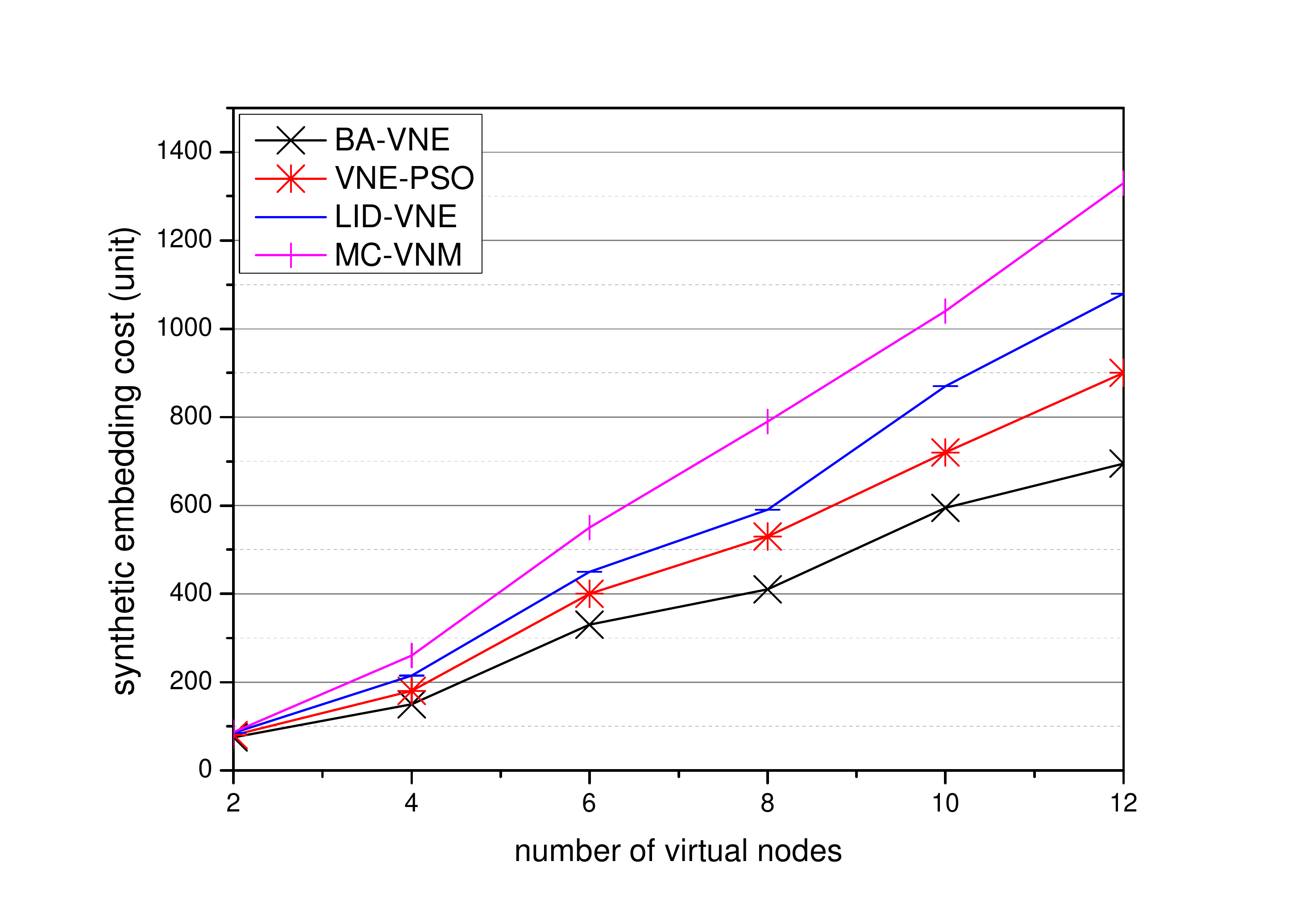}
	\caption{Synthetic embedding cost}
	\label{fig_8}
\end{figure}

In the test synthetic embedding cost experiment, we keep changing the number of virtual nodes. Change the 4 virtual nodes we set up earlier to 2, 4, 6, 8, 10, and 12 virtual nodes. As the number of virtual nodes increases, the total cost of various embedding algorithms will inevitably increase. The BA-VNE algorithm is optimal. It can be seen from the Fig. \ref{fig_8} that the algorithm has the lowest comprehensive embedding cost in the case of various virtual nodes. Then comes the VNE-PSO algorithm, LID-VNE algorithm and the MC-VNM algorithm with the highest comprehensive embedding cost.

\textit{Analysis of experimental results:} BA-VNE algorithm not only selects the maximum bandwidth link for mapping, but also takes into account the impact of mapping cost on the performance of the algorithm. Before mapping virtual nodes, the cost of mapping is estimated. By sorting the candidate nodes, the substrate nodes with the lowest estimated mapping cost are selected for embedding. This can reduce the mapping cost to some extent. VNE-PSO algorithm is worse than that of BA-VNE algorithm. Although the algorithm considers the cost estimation of node and link embedding, it selects the boundary node from the node closest to the boundary node. So this algorithm doesn't make a very good decision. LID-VNE is based on the distributed VNE architecture. Due to the business competition among SPs, they don't intend to disclose the details of network topology in their respective domains. Therefore, VNRs can only be embedded by randomly selecting substrate nodes and links, which leads to high mapping cost. The reason why the total mapping cost of MC-VNM algorithm is the highest is that the algorithm doesn't use the traditional natural heuristic algorithm to solve the problem of VNE. The natural heuristic algorithm has been proved to be an effective method to solve NP-hard problems. MC-VNM algorithm simply uses the greedy strategy to select nodes and links to map, so the effect is the worst.

\textbf{Experiment 4:} VNR acceptance rate comparison test.

\begin{figure}[!h]
	\includegraphics[width=1.0\columnwidth]{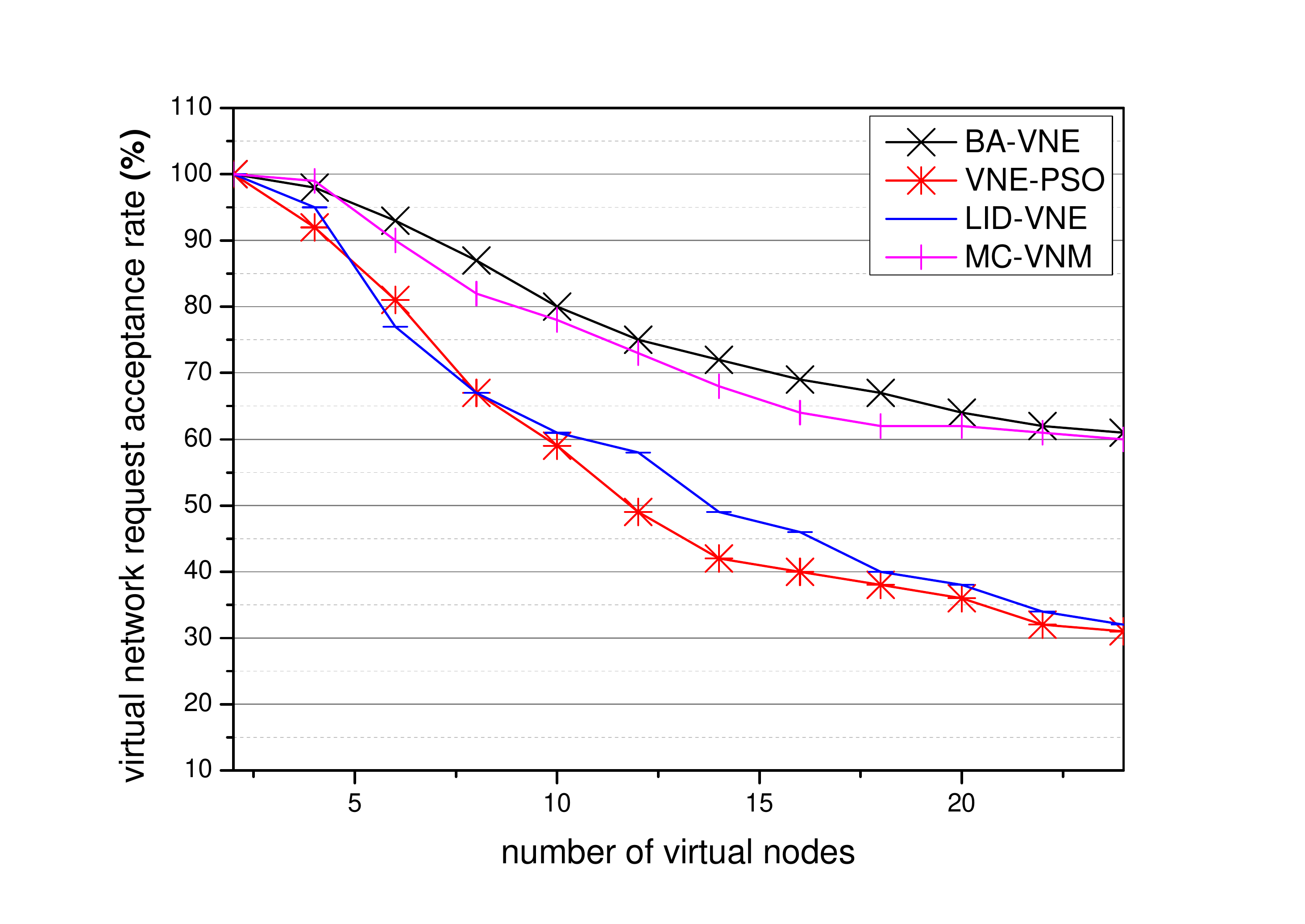}
	\caption{Virtual network request acceptance rate}
	\label{fig_9}
\end{figure}

With the increase of the number of virtual nodes, the acceptance rate of VNE requests of the four algorithms presents a downward trend. However, the virtual request acceptance rate of BA-VNE and MC-VNM is obviously better than that of VNE-PSO and LID-VNE. The experimental results are shown in Fig. \ref{fig_9}. In the MC-VNM and BA-VNE algorithms, the receiving rate of VNRs will stabilize around 60\%, while VNE-PSO and LID-VNE algorithms will drop to around 30\%.

\textit{Analysis of experimental results:} With the continuous consumption of the underlying network resources, the acceptance rate of VNRs of all the algorithms is constantly decreasing. The VNR acceptance rate of the BA-VNE algorithm is higher than that of the other three algorithms, which benefits from the idea of load balancing while optimizing the bandwidth index. In order to as far as possible in the case of no loss of bandwidth index, but also as good as possible reception rate. The MC-VNM algorithm uses the Kruskal spanning tree idea to select the link-first embedding with the minimum link unit price. The node embedding scheme is determined by the link embedding scheme, which also has a high VNR acceptance rate. The reason why VNE-PSO algorithm and LID-VNE algorithm have low acceptance rate of VNRs is that the methods of selecting substrate nodes are too rigid. In the process of VNE, there are often insufficient node resources or link resources. Therefore, the performance of these two algorithms is poor.

\textbf{Experiment 5:} Average embedding delay comparison experiment.

\begin{figure}[!h]
	\includegraphics[width=1.0\columnwidth]{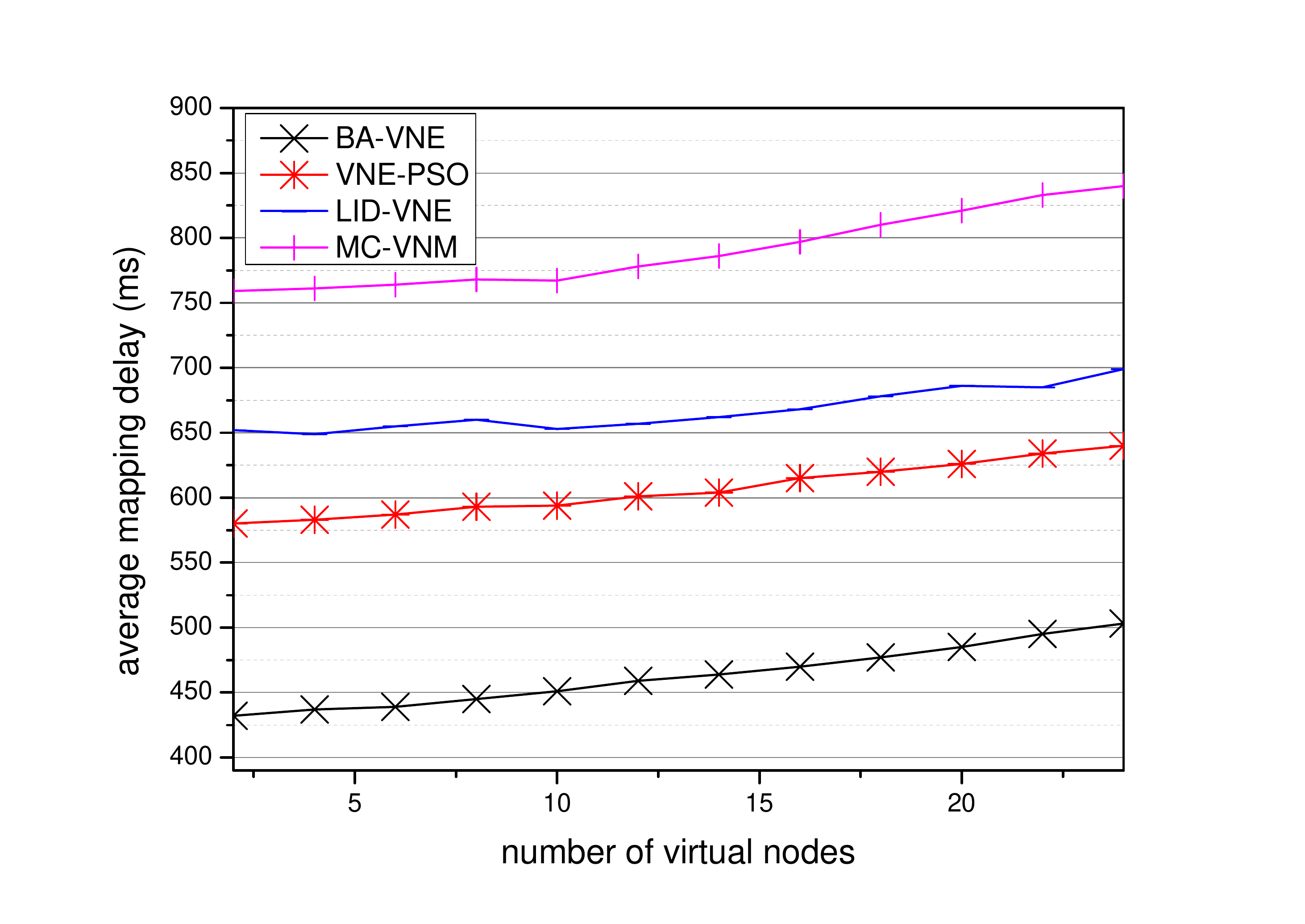}
	\caption{average embedding delay}
	\label{fig_10}
\end{figure}

The experimental results of mean embedding delay are shown in Fig. \ref{fig_10}. Taking the number of nodes requested by the virtual network as the independent variable, the average embedding delay of the four algorithms increases with the number of virtual nodes. The average delay of MC-VNM algorithm is the largest, reaching above 850. Next up are LID-VNE algorithm and VNE-PSO algorithm, both with a delay of 600 or more. Our BA-VNE algorithm has the lowest latency, below 500.

\textit{Analysis of experimental results:} The difference of delay is mainly reflected in the choice of link mapping. First of all, LID-VNE algorithm adopts distributed VNE architecture. The disadvantages of using this architecture have been mentioned in the analysis of Experiment 3. VNE-PSO algorithm only selects boundary nodes as candidate nodes, which limits the possibility of selecting other substrate links. The virtual link can only select the substrate link connected with the boundary node for mapping and a single link selection may produce a large mapping delay. BA-VNE algorithm selects the link with the largest bandwidth for mapping and uses PSO algorithm for optimization. Experimental results show that this method has achieved good results. As shown in Fig. \ref{fig_10}, the experimental effect of MC-VNM algorithm is the worst. Different from the other three algorithms, this algorithm does not use efficient heuristic method as a solution. It uses a relatively inefficient greedy strategy to complete the VNE problem, so the experimental results are poor.

\textbf{Experiment 6:} Link utilization comparison experiment.

\begin{figure}[!h]
\includegraphics[width=1.0\columnwidth]{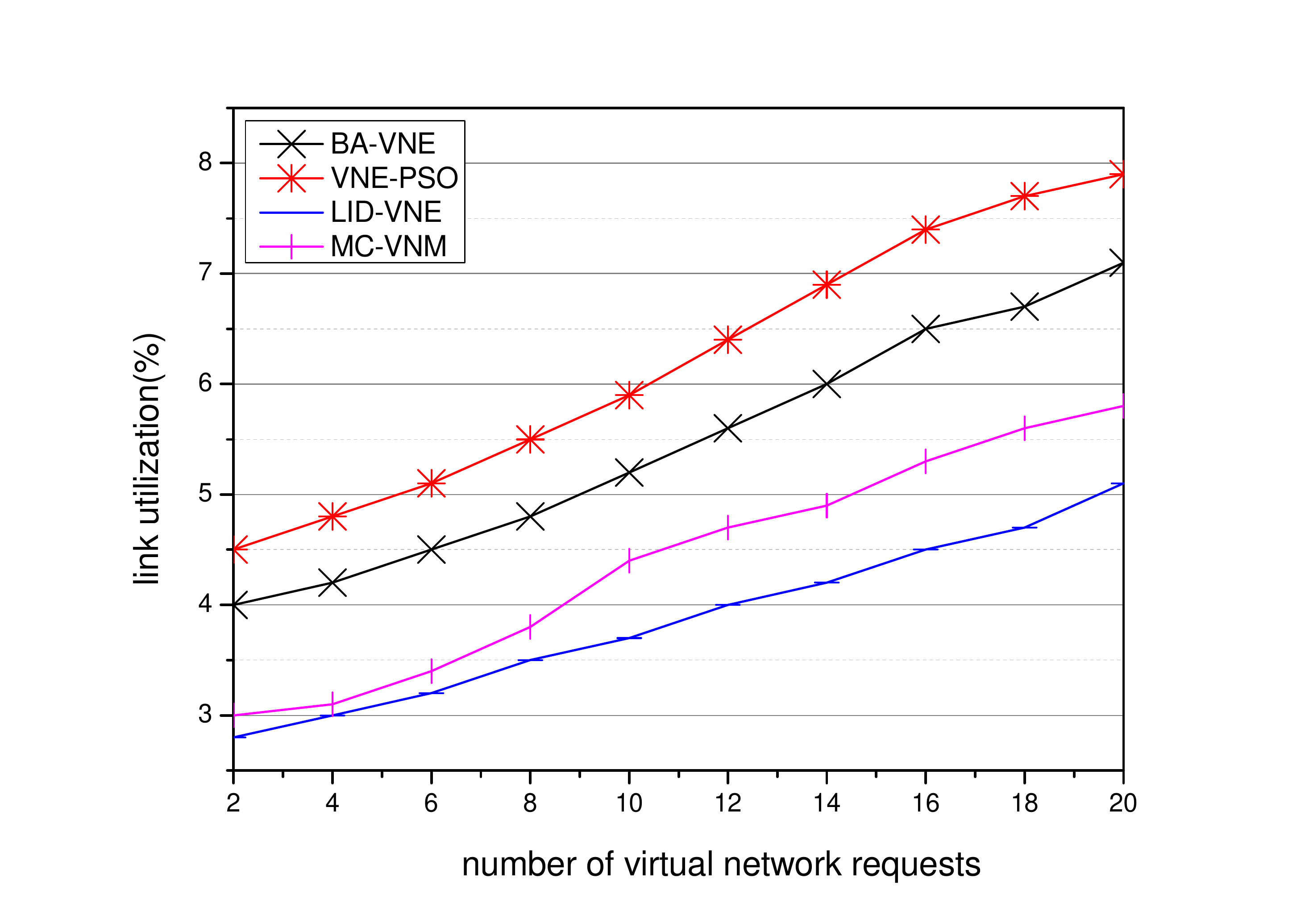}
\caption{Link utilization}
\label{fig_11}
\end{figure}

Link utilization is an important index to evaluate the performance of VNE algorithm. The experiment we designed takes the number of VNRs as an independent variable and the experimental results are shown in Fig. \ref{fig_11}. With the increase of the number of VNE requests, the link utilization of various algorithms increases. VNE-PSO algorithm has the highest link utilization rate, which can reach about 8\%. The second is BA-VNE algorithm and MC-VNM algorithm. The lowest link utilization is the LID-VNE algorithm, which is only 5\%.

\textit{Analysis of experimental results:} The difference of link utilization is caused by the link embedding strategy. VNE-PSO, BA-VNE and MC-VNM all adopt the shortest path method to embed virtual links, while LID-VNE adopts the multi-commodity flow method to embed virtual links. Experimental results show that the algorithm using shortest path strategy for link embedding has higher average link utilization. Because BA-VNE algorithm only selects links with larger link bandwidth for embedding, the range of embedded links is reduced, but VNE-PSO does not pay attention to this point. Therefore, the link utilization of BA-VNE algorithm is lower than that of VNE-PSO algorithm.

\section{Conclusion}

Intelligent application in wireless network communication mode brings great pressure and challenge to traditional network architecture. Especially in the aspect of network resource scheduling, how to use wireless network to achieve efficient and reasonable resource scheduling is an important topic. Based on the study of NV and multi domain VNE, this paper proposes a BA-VNE algorithm. This algorithm focuses on the bandwidth constraint, introducing mutation factor into traditional PSO can effectively avoid falling into the local optimal situation. Finally, BA-VNE algorithm is compared with MP-VNE, VNE-PSO, LID-VNE and MC-VNM algorithms. The simulation results show that the multi-domain VNE scheme proposed in this paper cannot only select the substrate link with the most remaining bandwidth resources for embedding. Moreover, it also performs well in terms of embedding cost, VNR acceptance rate and other aspects while ensuring the maximum bandwidth. Therefore, the VNE scheme with the largest bandwidth can solve the practical problems such as the IoV, video data backup and so on. It has good application significance.

\ifCLASSOPTIONcaptionsoff
  \newpage
\fi

\section*{Biographies}

\begin{IEEEbiography}[{\includegraphics[width=1in,height=1.25in,clip,keepaspectratio]{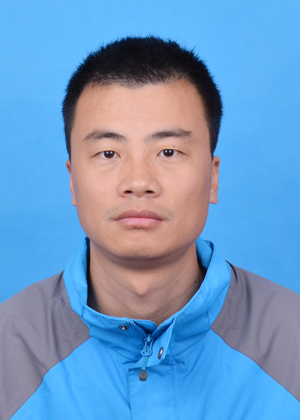}}]{Peiying Zhang}
is currently an Associate Professor with the College of Computer Science and Technology, China University of Petroleum (East China). He received his Ph.D. in the School of Information and Communication Engineering at the Beijing University of Posts and Telecommunications in 2019. His research interests include semantic computing, future internet architecture, network virtualization, and artificial intelligence for networking. He has published more than 50 papers in prestigious journals and conferences.
\end{IEEEbiography}

\begin{IEEEbiography}[{\includegraphics[width=1in,height=1.25in,clip,keepaspectratio]{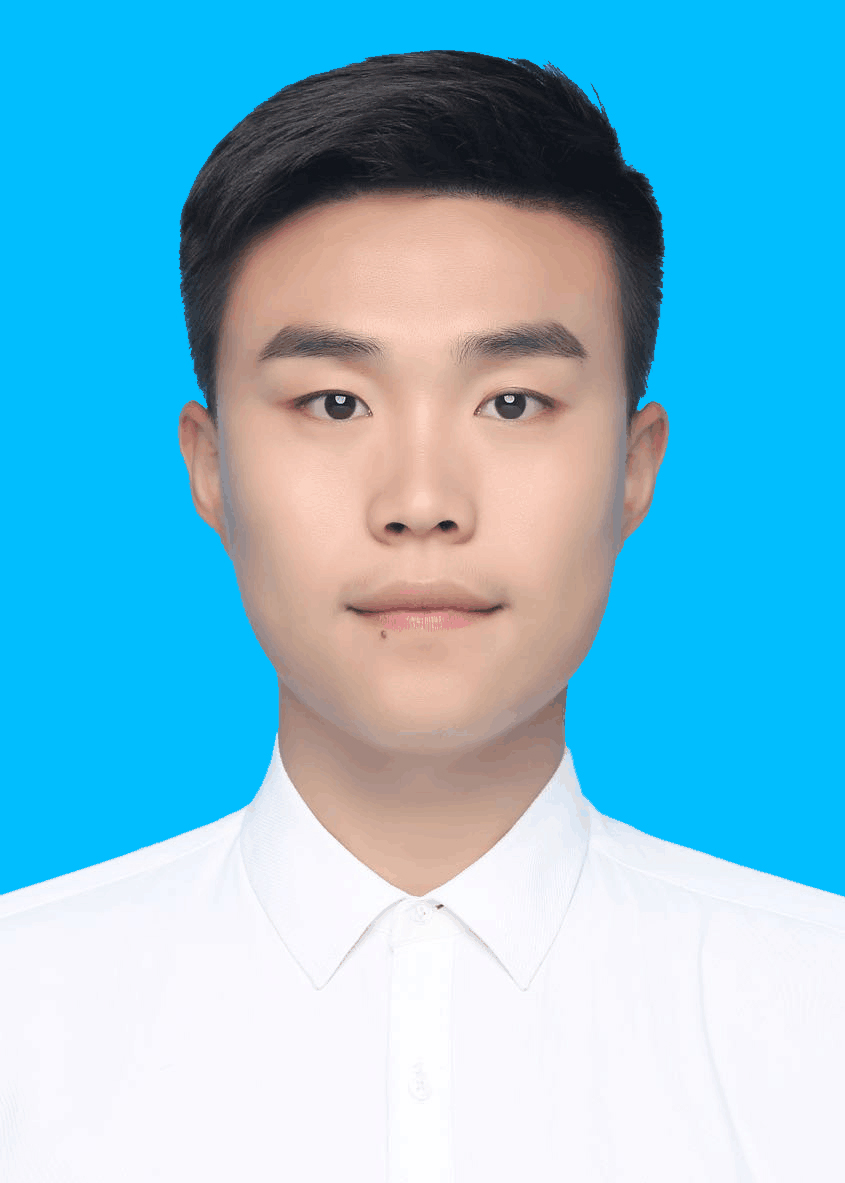}}]{Chao Wang}
is a graduate student in the College of Computer Science and Technology, China University of Petroleum (East China). His research interests include network artificial intelligence and network virtualization.
\end{IEEEbiography}

\begin{IEEEbiography}[{\includegraphics[width=1in,height=1.25in,clip,keepaspectratio]{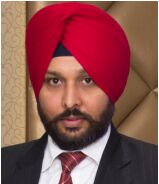}}]{Gagangeet Singh Aujla} received his Ph.D. in Computer Science and Engineering from Thapar Institute of Engineering and Technology, Patiala, Punjab, India in 2018. He received the B.Tech and M.Tech. degrees in Computer Science and Engineering from Punjab Technical University, Jalandhar, Punjab, India, in 2003 and 2013, respectively. Currently, he is working as an PostDoc Research Associate at the School of Computing, Newcastle University, United Kingdom. He is also an Associate Professor in Computer Science and Engineering Department, Chandigarh University, Mohali, Punjab, India. Prior to this, he was working as a Research Associate in Indo-Austria Research project sponsored by Department of Science and Technology, Government of India and Ministry of Science, Austria. He also worked as a Project fellow in Haryana State Center of Science and Technology funded research project on Smart Grid. He is recipient of 2018 IEEE TCSC outstanding PhD dissertation award. He is on the Editorial Board of the Sensors Journal. He has guest edited a number of Special Issues in IEEE Transactions on Industrial Informatics, IEEE Network, Computer Communications, Software: Practice and Experience, and Transactions on Emerging Telecommunications Technologies. He is Senior Member of the IEEE and Member of the ACM.
\end{IEEEbiography}

\begin{IEEEbiography}[{\includegraphics[width=1in,height=1.25in,clip,keepaspectratio]{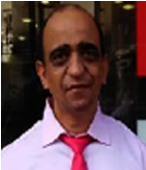}}]{Neeraj Kumar} received his Ph.D. in CSE from Shri Mata Vaishno Devi University, Katra (Jammu and Kashmir), India in 2009, and was a postdoctoral research fellow in Coventry University, Coventry, UK. India. He is working as a Full Professor in the Department of Computer Science and Engineering, Thapar Institute of Engineering and Technology, Patiala, Punjab, India. He is an Adjunct Professor at the King Abdulaziz University, Jeddah, Saudi Arabia and also at the Asia University, Taiwan. He is visiting research fellow at Coventry University, UK. He is a Technical Editor of IEEE Network Magazine, and IEEE Communication Magazine. He is an Associate Editor of IEEE Transactions on Sustainable Computing, ACM Computing Surveys, JNCA, Elsevier, IJCS, Wiley, and Security and Communication, Wiley and on the Editorial Board of Computer Communications, Elsevier. He is senior member of the IEEE. He has more than 8800 citations to his credit with current h-index of 51. He has won the best papers award from IEEE Systems Journal and ICC 2018, Kansas city in 2018. He has edited more than 10 journals special issues of repute and published four books from CRC, Springer, IET UK, and BPB publications. He is Senior Member of the IEEE.
\end{IEEEbiography}

\begin{IEEEbiography}[{\includegraphics[width=1in,height=1.25in,clip,keepaspectratio]{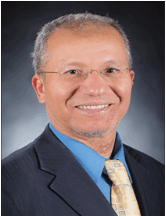}}]{Mohsen Guizani} received his B.S. (with distinction) and M.S. degrees in electrical engineering, and M.S. and Ph.D. degrees in computer engineering from Syracuse University, New York, in 1984, 1986, 1987, and 1990, respectively. He is currently a professor in the Computer Science and Engineering Department at Qatar University. Previously, he served in different academic and administrative positions at the University of Idaho, Western Michigan University, the University of West Florida, the University of Missouri-Kansas City, the University of Colorado-Boulder, and Syracuse University. His research interests include wireless communications and mobile computing, computer networks, mobile cloud computing, security, and smart grid. He is currently the Editor-in-Chief of IEEE Network, serves on the Editorial Boards of several international technical journals, and is the Founder and Editor-in-Chief of the Wireless Communications and Mobile Computing journal (Wiley). He is the author of nine books and more than 500 publications in refereed journals and conferences. He has guest edited a number of Special Issues in IEEE journals and magazines. He has also served as a TPC member, Chair, and General Chair of a number of international conferences. Throughout his career, he received three teaching awards and four research awards. He also received the 2017 IEEE Communications Society WTC Recognition Award as well as the 2018 AdHoc Technical Committee Recognition Award for his contribution to outstanding research in wireless communications and ad hoc sensor networks. He was the Chair of the IEEE Communications Society Wireless Technical Committee and the Chair of the TAOS Technical Committee. He served as a IEEE Computer Society Distinguished Speaker and is currently an IEEE ComSoc Distinguished Lecturer. He is Fellow of the IEEE and a Senior Member of ACM.
\end{IEEEbiography}

\end{document}